# Embedded Automotive System Development Process
## Steer-By-Wire System

Joachim Langenwalter
jlangenwalter@mathworks.com
The MathWorks

*Abstract*

Model based design enables the automatic generation of final-build software from models for high-volume automotive embedded systems.
This paper presents a framework of processes, methods and tools for the design of automotive embedded systems. A steer-by-wire system serves as an example.

*Process*

Model-based design supports the needs of controls/DSP systems engineers and software developers by providing a common environment for graphical specification and analysis. In this process, models are made and used to specify system data, interfaces, feedback control logic, discrete/state logic, and real-time behavior.

A common way to view a software process is through use of the V diagram. The diagram corresponds to most engineering processes, however, the process is iterative with many repetitive steps throughout the development life cycle.

The software process in this diagram is comprised of the following:
- **Development** (Requirements, Design, Coding, Integration)
- **Verification and Validation** (V&V)
- **Integral** (Software Configuration Management, Requirements Traceability and Documentation)

*Methods & Tools*

Model-based design methods are employed during the software engineering process.

The development methods include:
1. Behavioral Modeling
2. Detailed Software Design
3. Distributed Architecture Design
4. Production Code Generation
5. Embedded Target Integration

The V&V methods include:
1. Simulation and Analysis
2. Rapid Prototyping
3. Model Testing and Coverage
4. Code Tracing and Reviews
5. Hardware-In-the-Loop (HIL) Testing

The Integral methods include:
1. Source Control Interface
2. Requirements Management Interface
3. Report Generation

*Behavioral Modeling*

Models are used for specifying requirements and design for all aspects of every individual subsystem (e.g. steer-by-wire in Figure 1 and 2).

A typical system includes:
- Inputs (e.g., steering wheel sensors)
- Controller or DSP Model
- Plant Model (DC motor, rack and pinion, wheels )
- Outputs (change of direction)

As shown, a system model can be created to represent the desired behavior using control system block diagrams for feedback control, state machines for discrete events and conditional logic, and DSP blocks for filters.

*Simulation and Analysis*

The model is executed and analyzed to ensure that the requirements are satisfied, using methods such as time- or event-based simulation and frequency domain analysis.

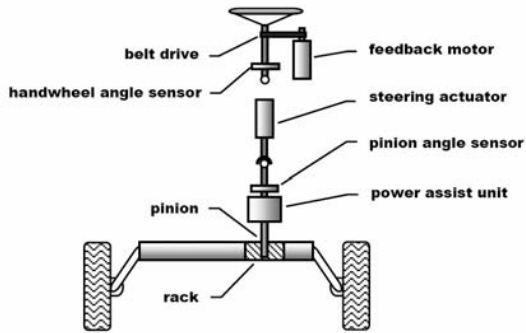

**Figure 1: Steer-By-Wire system**

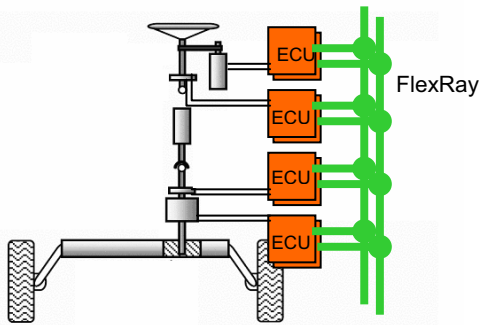

**Figure 2: Steer-By-Wire system with fault tolerant redundant bus system (FlexRay)**

*Rapid Prototyping*

Due to inaccurate plant models and insufficient processing power rapid prototyping is highly useful because it replaces the plant model with the physical plant.

*Detailed Software Design*

With model-based design, the same model used for algorithm specification and validation is refined and constrained by the software engineers as part of the production code generation process.

*Model Testing*

It is more beneficial to test the model on a desktop rather than deploy it on hardware for build and integration. Source code-based testing has existed for many years, but recent methods now allow for model testing and structural coverage.

| Model Hierarchy/Complexity: | | Test 1 | | |
|---|---|---|---|---|
| | | D1 | C1 | MCDC |
| 1. req_test | 35 | 84% | 70% | 50% |
| 2. .... Logic | 25 | 78% | 75% | 50% |
| 3. ....... SF: Logic | 24 | 78% | 75% | 50% |
| 4. ......... SF: Altitude | 11 | 100% | 83% | 67% |
| 5. ........... SF: Active | 4 | 100% | NA | NA |
| 6. ......... SF: GS | 13 | 61% | 67% | 33% |
| 7. ........... SF: Active | 6 | 50% | NA | NA |
| 8. ............. SF: Coupled | 3 | 33% | NA | NA |

**Figure 3: Coverage for power management design in Fig. 3**

*Production Code Generation*

Now that the model has been verified and validated, it is time to generate code automatically from the model.

*Hardware-In-the-Loop Testing*

Once the controller has been built, a series of open- and closed-loop tests can be performed with the real-time plant model in the loop.

*Integral Components*

Most software standards require traceability of requirements, perhaps originating in other requirements tools, throughout development.

*Conclusion*

Major software evolutions occur when the full software engineering process activities are supported. Improving bits and pieces alone is insufficient.